\documentclass[journal,10pt,twocolumn]{IEEEtran}
\IEEEoverridecommandlockouts
\usepackage{amsmath,amssymb,amsfonts,float,graphicx,steinmetz,bm,subfigure,physics,mathtools,url,color,xcolor,etoolbox,ifthen,cite,physics}
\usepackage{textcomp}
\usepackage{tikz}
\usepackage[linesnumbered,ruled,vlined]{algorithm2e}
  
\SetKwInput{KwInput}{Input}                
\SetKwInput{KwOutput}{Output}

\definecolor{lime}{HTML}{A6CE39}
\DeclareRobustCommand{\orcidicon}{%
	\begin{tikzpicture}
	\draw[lime, fill=lime] (0,0) 
	circle [radius=0.16] 
	node[white] {{\fontfamily{qag}\selectfont \tiny ID}};
	\draw[white, fill=white] (-0.0625,0.095) 
	circle [radius=0.007];
	\end{tikzpicture}
	\hspace{-2mm}
}

\foreach \x in {A, ..., Z}{%
	\expandafter\xdef\csname orcid\x\endcsname{\noexpand\href{https://orcid.org/\csname orcidauthor\x\endcsname}{\noexpand\orcidicon}}
}

\begin{document}
\title{ 
Machine Learning Based Parameter Estimation of Gaussian Quantum States
\thanks{The work of Neel Kanth Kundu and Matthew R. McKay was supported by the Hong Kong Research Grants Council (grant number C6012-20G). The work of Ranjan K. Mallik was supported in part by the Science and Engineering Research Board, a Statutory Body of the Department of Science and Technology, Government of India, under the J. C. Bose Fellowship.}
\vspace{-0.1in}
}

\author{
Neel Kanth Kundu\orcidA{}, {\em Graduate Student Member, IEEE},
Matthew R. McKay\orcidB{}, {\em Fellow, IEEE}, \\ and Ranjan K. Mallik\orcidD{}, {\em Fellow, IEEE}
\thanks{
N. K. Kundu and M. R. McKay are with the Department of
Electronic and Computer Engineering, The Hong Kong University of
Science and Technology, Clear Water Bay, Kowloon, Hong Kong
(e-mail: nkkundu@connect.ust.hk,  m.mckay@ust.hk).

R. K. Mallik is with the Department of Electrical Engineering, Indian Institute of Technology Delhi, New Delhi, India (e-mail: rkmallik@ee.iitd.ernet.in). }

}


\markboth{  This work has been submitted to the IEEE for possible publication. Copyright may be transferred without notice.}{This work has been submitted to the IEEE for possible publication. Copyright may be transferred without notice.}


\maketitle
\begin{abstract}
We propose a machine learning framework for parameter estimation of single mode Gaussian quantum states. Under a Bayesian framework, our approach estimates parameters of suitable prior distributions from measured data. For phase-space displacement and squeezing parameter estimation, this is achieved by introducing Expectation-Maximization (EM) based algorithms, while for phase parameter estimation an empirical Bayes method is applied. The estimated prior distribution parameters along with the observed data are used for finding the optimal Bayesian estimate of the unknown displacement, squeezing and phase parameters. Our simulation results show that the proposed algorithms have estimation performance that is very close to that of ‘Genie Aided’ Bayesian estimators, that assume perfect knowledge of the prior parameters. Our proposed methods can be utilized by experimentalists to find the optimum Bayesian estimate of parameters of Gaussian quantum states by using only the observed measurements without requiring any knowledge about the prior distribution parameters.
\end{abstract}


\section{Introduction} \label{sec:introduction}
Quantum metrology is one of the important quantum technologies that uses quantum mechanics to study the ultimate limits with which physical quantities can be estimated \cite{giovannetti2006quantum,giovannetti2011advances}. Quantum estimation has been well studied over the past few decades to understand the ultimate limits of parameter estimation achievable by quantum measurements\cite{paris2009quantum,vsafranek2018estimation,martinez2017quantum,pinel2013quantum}. In quantum parameter estimation, first an initial probe state $\rho_\text{ in}$ is prepared which undergoes a transformation $\mathcal{E}_{\theta}(\cdot)$, that encodes the unknown parameter $\theta$ to the state $\rho_{\theta}=\mathcal{E}_{\theta}(\rho_\text{ in})$ \cite{jiang2014quantum,morelli2021bayesian,pinel2013quantum}. This encoded quantum state $\rho_{\theta}$ is then measured to estimate the unknown parameter $\theta$. In order to estimate the parameter, first the optimal positive operator valued measure (POVM) operator is constructed that can extract the maximum information about $\theta$ from the encoded state. Then, classical processing of the measurement outcome from the POVM is carried out to find the optimal estimate of $\theta$. Quantum estimation theory provides a lower bound on the variance of the unbiased estimate $\theta_{\text{ est}}$ in the form of the quantum Cram{\'e}r Rao bound (QCRB) \cite{paris2009quantum}. This bound arises due to quantum uncertainty and it depends on the quantum Fisher information (QFI) \cite{paris2009quantum}. Previous works have extensively studied the QCRB and determined the QFI to theoretically lower bound the variance of the estimated parameter for different quantum states, including single mode and multimode Gaussian coherent states, squeezed states, $N00N$ states and single photon states \cite{pinel2013quantum,vsafranek2018estimation,pezze2018quantum,gorecki2020pi,jarzyna2015true}. The QCRB can be achieved asymptotically when a large number of measurements on independent probe states are available. This requires a `frequentist' estimation approach such that the QCRB can be achieved asymptotically \cite{pinel2013quantum,monras2013phase,jiang2014quantum,vsafranek2018estimation}. 

However, in practice it is desirable to estimate the parameter from a limited number of measurements. In the limited observation scenario, the estimation accuracy can be improved by incorporating prior information about the unknown parameter. Bayesian estimation scheme uses the prior belief about the parameter and updates its belief as observations become available. In the frequentist approach the unknown parameters are treated as fixed whereas in the Bayesian approach they are treated as random variables with an associated prior distribution function. The application of the Bayesian estimation scheme for various quantum parameter estimation problems has previously been studied in \cite{morelli2021bayesian,martinez2017quantum,rubio2020bayesian,rubio2019quantum}. The Bayesian estimation approach utilizes knowledge of the prior distribution of the unknown parameter $\theta$ and the likelihood of the observed data to find the posterior distribution of $\theta$. The optimal Bayesian estimate of $\theta$ is given by the posterior mean, and the variance of the estimate is given by the posterior variance. In order to determine the Bayesian estimate, the prior distribution parameters should be known accurately, which is difficult to obtain in practice. Thus, efficient and practically realizable methods are required for Bayesian parameter estimation of quantum states. 

In this work we propose a machine learning based method to find the optimal Bayesian estimate of $\theta$. Our proposed method does not require knowledge of the prior distribution parameters, as these are estimated from the measured data. Machine learning has the ability to extract useful information and patterns from data which has led to its successful applications in computer vision, natural language processing, classification, recommendation systems, etc. \cite{bishop2006pattern}. Recently researchers have extensively investigated the application of machine learning for quantum information processing tasks like quantum channel estimation, Hamiltonian estimation, quantum state discrimination and quantum many-body physics \cite{carleo2019machine,ismail2019integrating,xu2019generalizable,greplova2017quantum,nolan2020machine,dunjko2018machine,xiao2019continuous,bertalan2019learning,wang2016discovering}. Motivated by these studies, in this work we propose a machine learning based quantum parameter estimation scheme.

We consider the problem of parameter estimation for continuous-variable (CV) single-mode Gaussian quantum states. CV Gaussian quantum states play an important role in quantum key distribution, quantum computation, quantum metrology and quantum secure direct communications \cite{weedbrook2012gaussian,grosshans2002continuous,ralph1999continuous,guerra2016quantum,srikara2020continuous,chai2019novel,yuan2015continuous}. Gaussian quantum states can be easily generated in the laboratory and are easy to analyse since their Wigner functions admit a Gaussian form \cite{weedbrook2012gaussian}. We propose machine learning based methods for three important estimation problems for Gaussian states: (i) phase-space displacement estimation, (ii) single-mode squeezing parameter estimation, and (iii) phase estimation. This problem has recently been studied in \cite{morelli2021bayesian}, where authors have proposed Bayesian estimation schemes. However, the method in \cite{morelli2021bayesian} requires perfect knowledge of the prior distribution parameters, and no method has been proposed to obtain these prior parameters in practice. Since the performance of the Bayesian estimate depends on the knowledge of the prior distribution parameters \cite{fortunati2017performance}, it is important to find a practical estimation scheme that can estimate the prior parameters along with the optimal Bayesian estimate of the quantum parameter of interest. Our work addresses this problem. Moreover, in contrast to the previous work \cite{morelli2021bayesian} that assumed a single measurement scenario, here we consider a scenario where multiple independent measurement outcomes are jointly processed to find the optimum estimate of the unknown parameter. This multiple measurement scenario has also been considered by recent works on data driven quantum information processing, where authors have proposed machine learning based Hamiltonian estimation \cite{bertalan2019learning}, single electron Rabi frequency estimation \cite{greplova2017quantum}, single qubit rotation estimation \cite{nolan2020machine}, and machine learning based quantum interferometry \cite{xiao2019continuous}. 

Here, we focus on single mode Gaussian quantum states and propose a machine learning based method to estimate the prior distribution parameters from the observed measurement data. The estimated prior parameters are used to find the optimal Bayesian estimate of the parameters. We assume a conjugate prior for the displacement, squeezing and phase parameters which permits analytical characterization of the posterior distribution of the parameter. The prior parameters are estimated by maximizing the log-likelihood of the observed data. We propose an Expectation-Maximization (EM) algorithm for estimating the prior parameters of the displacement and squeezing parameters, along with an empirical Bayes method for estimating the prior for the phase parameter. The estimated prior parameters are used to find the optimal Bayesian estimate of the displacement, squeezing and phase parameter. Our simulation results show that the proposed algorithms have estimation performance that is very close to that of ‘Genie Aided’ Bayesian estimators, that assume perfect knowledge of the prior parameters. Thus, the proposed methods can be utilized by experimentalists to find the optimum Bayesian estimate of parameters by using only the observed measurements without requiring knowledge about the prior distribution parameters.

The paper is organized as follows. In Section \ref{CV_review}, we briefly review some important concepts related to continuous variable Gaussian quantum states. In Section \ref{displacement}, we present the machine learning based displacement estimation for coherent Gaussian states. Machine-learning-based squeezing estimation and phase estimation of Gaussian quantum states are presented respectively in Sections \ref{squeezing} and \ref{phase}. Finally some concluding remarks are made in Section \ref{conclusion}.

\section{Formulation} \label{CV_review}

\subsection{Continuous Variable States}

In this section we briefly present the formalism of CV Gaussian quantum states used in this paper. For a detailed introduction to Gaussian quantum information, readers are referred to prior comprehensive survey papers \cite{braunstein2005quantum,weedbrook2012gaussian,ferraro2005gaussian}. 

Multi-mode optical fields can be represented as a collection of bosonic modes. CV systems consist of $N$ bosonic modes, i.e., $N$ harmonic oscillators. We index each mode by $k$, and $\hat{a}_k,\hat{a}_k^{\dagger}$ denote the annihilation and creation operators of the $k$-th mode respectively. These modes allow us to construct the quadrature operators for each mode, $\hat{q}_k=(\hat{a}_k^{\dagger}+\hat{a}_k)/\sqrt{2}$ and $\hat{p}_k=i(\hat{a}_k^{\dagger}-\hat{a}_k)/\sqrt{2}$ (e.g., see \cite{weedbrook2012gaussian}). These operators provide position and momentum observables and have continuous spectra and eigenstates, $\{\ket{q}\}_{q\in\mathbb{R}}$ and $\{\ket{p}\}_{p\in\mathbb{R}}$. Below, we write the quadrature operators as a single vector $\hat{\mathbf{x}}=(\hat{q}_1,\hat{p}_1,\ldots,\hat{q}_N,\hat{p}_N)^T$. The $N$ bosonic modes are associated with a tensor product Hilbert space $ \mathcal{H}^{\otimes N} = \otimes_{k=1}^{N}\mathcal{H}_k $, where $\mathcal{H}_k $ is the infinite dimensional Hilbert space corresponding to the $k$-th mode.
Letting $\mathcal{D}(\mathcal{H}^{\otimes N}) $  denote the space of density operators, any state of a $N$-mode bosonic system is represented by a density operator $\rho\in\mathcal{D}(\mathcal{H}^{\otimes N})$, which is a unit trace and positive semi-definite operator \cite{weedbrook2012gaussian}. In CV systems, the quantum state can also be represented by a Wigner function $W(\mathbf{x})$~\cite{wigner1932quantum}. The transformation from the density matrix to the Wigner function is performed as follows: the characteristic function is defined by $\chi(\mathbf{x})=\Tr[\rho D(\mathbf{x})]$, where $D(\mathbf{x})=\exp(i\sqrt{2} \mathbf{x}^T \boldsymbol{\Omega} \mathbf{x})$ is the Weyl displacement operator and $\boldsymbol{\Omega}$ is the symplectic matrix~\cite{arvind1995real}, and then the Wigner function $W(\mathbf{x})$ is given by the Fourier transform of the characteristic function $\chi(\mathbf{x})$.

\subsection{Gaussian States}

CV states for which the Wigner function $W(\mathbf{x})$ admits a multivariate Gaussian form
\begin{equation}
    W(\mathbf{x}) = \frac{\exp[-\frac{1}{2}(\mathbf{x}-\mathbf{\Bar{x}})^T \mathbf{V}^{-1}(\mathbf{x}-\mathbf{\Bar{x}})]}{(2\pi)^N \sqrt{\text{det}\left(\mathbf{V} \right)}}
    \label{sec2.3}
\end{equation}
are called Gaussian states. These states are analytically tractable since they are characterized only with the mean vector $\mathbf{\Bar{x}} = \text{ Tr}\left(\hat{\mathbf{x}}\rho \right)$ and the covariance matrix $\mathbf{V}$, which is a $2N\times 2N$ real symmetric matrix with entries 
\begin{equation}
    \mathbf{V}_{i,j} = \frac{1}{2} \left\langle\left\{\hat{\mathbf{x}}_{i}-\left\langle\hat{\mathbf{x}}_{i}\right\rangle, \hat{\mathbf{x}}_{j}-\left\langle\hat{\mathbf{x}}_{j}\right\rangle\right\}\right\rangle,
    \label{sec2.4}
\end{equation}
where $\langle \cdot \rangle$ is the expectation operator and $\{,\}$ is the anti-commutator operator. This property of Gaussian states enables one to compactly treat the states which expand in an infinite-dimensional Hilbert space with a finite number of degrees of freedom. A large class of quantum states including vacuum states, thermal states of black-body radiation, coherent states of laser radiation, and squeezed light fall into the category of Gaussian quantum states \cite{weedbrook2012gaussian}.

In this paper, we focus only on single-mode Gaussian quantum states since they are quintessential building blocks for quantum information processing tasks like CV quantum cryptography \cite{weedbrook2012gaussian}, and can be easily generated and manipulated in the laboratory. Any of such states are generated by displacing, squeezing, and rotating states. We introduce each of these operations below. First, displacement is performed by 
\begin{equation}
  \hat{D}(\alpha) \coloneqq \exp\left[\alpha \hat{a}^{\dagger}-\alpha^{*}\hat{a} \right]
  \label{sec2.5}
\end{equation}
with $\alpha \in \mathbb{C}$ being the complex amplitude. A coherent state is created by displacing the vacuum state $\ket{0}$, i.e., $\ket{\alpha}= \hat{D}(\alpha)\ket{0}$. 
The mean vector of such a state is determined by the displacement parameter and given by $\Bar{\mathbf{x}}=\sqrt{2}\left[ \text{ Re}(\alpha), \text{ Im}(\alpha) \right]^T$.

Squeezed states are a wider class of quantum states that allow for different variances for the two quadratures. The squeezing operation reduces the variance of either quadrature, while the variance of the other quadrature increases such that the Heisenberg uncertainty principle holds. The squeezing operator is defined as
\begin{equation}
    \hat{S}(r) \coloneqq \exp\left[  \frac{r}{2}\left(\hat{a}^2-\hat{a}^{\dagger 2} \right)\right],
    \label{sec2.7}
\end{equation}
where $r\in \mathbb{R}$ is the squeezing parameter. Positive (negative) values of $r$ perform squeezing along the $q$ ($p$) axis.

Lastly, we use the rotation operator defined as
\begin{equation}
    \hat{R}(\theta) \coloneqq \exp\left[ -i\theta \hat{a}^{\dagger} \hat{a} \right],
    \label{sec2.8}
\end{equation}
where $\theta \in [-\pi,\pi)$ is the rotation parameter. These three operators can produce any pure single-mode Gaussian state which is formulated as $\ket{\alpha,\theta,r}= \hat{D}(\alpha) \hat{R}(\theta) \hat{S}(r) \ket{0}$. The mean vector is given by $\bar{{\mathbf{x}}} = \sqrt{2}\left[ \text{ Re}(\alpha), \text{ Im}(\alpha) \right]^T $, and the covariance matrix by
\begin{equation}
    \mathbf{V} = \frac{1}{2} {\small \begin{bmatrix} \cosh{2r}-\cos{2 \theta} \sinh{2r} & \sin{2\theta} \sinh{2r} \\ \sin{2\theta} \sinh{2r} & \cosh{2r}+\cos{2 \theta} \sinh{2r}
    \end{bmatrix}.
    }
\end{equation}
Note that the order of the operators matters. If they are switched, the mean vector and the covariance matrix have different parameter dependencies. 

\subsection{Gaussian Measurements}

Any measurement of a quantum state can be represented by a POVM, i.e., a set of positive operators $\{E_{m}\}$ which sum to the identity. As we consider CV systems, we use continuous POVMs which gives a continuous set of operators and of measurement outcomes. Particularly, such operators whose measurement outcomes follow a Gaussian distribution on application to Gaussian quantum states are called Gaussian measurement operators. For instance, heterodyne and homodyne detection are the two most common Gaussian measurements. In heterodyne detection, both of the field quadratures are measured simultaneously, and the POVM is given by $ \left\{ \frac{1}{\pi} \ket{\beta} \bra{\beta} \right\}_{\beta \in \mathbb{C}}$ and composed of coherent states. In homodyne detection, only one of the mode quadratures, for example $\hat{q}$, is measured, and the POVM is given by the quadrature basis $\{\ket{q} \bra{q} \}_{q \in \mathbb{R}}$ and represented as projectors of the quadrature. 

In order to use the Gaussian quantum states as a resource in quantum information processing tasks, it is important to know the parameters of the Gaussian quantum state. Therefore, we now present different parameter estimation schemes for Gaussian quantum states which can be used by experimental physicists to estimate the parameters based on the measurement outcomes of the experiments.

\section{Displacement Estimation} \label{displacement}

We explore Bayesian estimation scenarios for the displacement by using Gaussian probe states and measurement. We consider that a displacement operator $\hat{D}(\alpha)$ is applied to the probe state described as a Gaussian state, and to estimate the displacement $\alpha$ we perform Gaussian measurements, which are repeated multiple times. We set our probe state as $\ket{\xi}=\hat{S}(\xi)\ket{0}$ with $\xi\geq0$ instead of the general Gaussian state $\hat{D}(\alpha')\hat{R}(\theta)\hat{S}(\xi)\ket{0}$, because Gaussian measurements are covariant under the action of displacement~\cite{chiribella2004covariant}. The unknown displacement parameter $\alpha$ can be complex, $\alpha=\Re[\alpha]+i\Im[\alpha]$. The Gaussian prior is the conjugate prior for a Gaussian likelihood of the data, which helps to analytically characterize the posterior distribution. We assume a Gaussian prior for both the real and imaginary part of $\alpha$, i.e.,
\begin{align}
    p(\alpha_{\mathrm{R}}) &= \frac{1}{ \sqrt{2 \pi \sigma_{0,\mathrm{R}}^2}} \exp\left(-\frac{(\alpha_{\mathrm{R}}-\alpha_{0,\mathrm{R}})^2}{2\sigma_{0,\mathrm{R}}^2}\right)\;, \nonumber \\
    p(\alpha_{\mathrm{I}}) &= \frac{1}{ \sqrt{2 \pi \sigma_{0,\mathrm{I}}^2}} \exp\left(-\frac{(\alpha_{\mathrm{I}}-\alpha_{0,\mathrm{I}})^2}{2\sigma_{0,\mathrm{I}}^2}\right) \;.
    \label{dis1}
\end{align}
The prior distribution parameters $\alpha_{0,\mathrm{R}}, \sigma_{0,\mathrm{R}}^2,  \alpha_{0,\mathrm{I}}, \sigma_{0,\mathrm{I}}^2 $ are estimated from the observed data.

\subsection{Heterodyne Measurement} \label{Appendix_hetero}
We employ the EM algorithm for displacement estimation using the heterodyne measurement scheme. The optimal POVM for heterodyne measurement is given by $\{\frac{1}{\pi} \ket{\beta} \bra{\beta} \}_{\beta \in \mathbb{C}}$, where the measurement outcome is Gaussian distributed. We consider a multiple measurement scenario where $M$ measurements are made and the outcomes are independent and identically distributed (i.i.d) with the probability to observe $\beta^{i} \in \mathbb{C}$ for the $i$-th measurement given by \cite[Eq.~18-19]{morelli2021bayesian}
\begin{equation}
    p(\beta_{R}^{i}|\alpha_R) = \frac{\sqrt{2}\exp\left[-\frac{2(\beta_{R}^{i}-\alpha_R)^2}{1+e^{-2r}} \right]}{\sqrt{\pi(1+e^{-2r})}}
    \label{Het1}
\end{equation}

\begin{equation}
    p(\beta_{I}^{i}|\alpha_I) = \frac{\sqrt{2}\exp\left[-\frac{2(\beta_{I}^{i}-\alpha_I)^2}{1+e^{2r}} \right]}{\sqrt{\pi(1+e^{2r})}}
    \label{Het2}
\end{equation}
where $r$ is the squeezing parameter and $\beta_{R}^{i} \, (\beta_{I}^{i} )$ represents the real (imaginary) part of the $i-$th measurement outcome $\beta^i$. Let $\bm{\beta}_R = [\beta_{R}^{1}, \beta_{R}^{2},\ldots, \beta_{R}^{M} ]^T$ and $\bm{\beta}_I = [\beta_{I}^{1}, \beta_{I}^{2},\ldots, \beta_{I}^{M} ]^T$, then the conditional likelihood of the observed data is given by
 \begin{align}
     p(\bm{\beta}_R|\alpha_R) &= \mathcal{N}\left(\alpha_R \bm{1}_M, \sigma^2_R \bm{I}_{M} \right) \nonumber \\ 
     p(\bm{\beta}_I|\alpha_I) &= \mathcal{N}\left(\alpha_I \bm{1}_M, \sigma^2_I \bm{I}_{M} \right)
 \label{Het3}
 \end{align}
where $\mathcal{N}\left(\bm{\mu}, \bm{\Sigma} \right)$ is the probability distribution function (pdf) of a Gaussian random vector with mean $\bm{\mu}$ and covariance matrix $\bm{\Sigma}$, and $\bm{1}_M = [1,1,\ldots,1]^T$. Further, the variance parameters $\sigma^2_R, \sigma^2_I$ are given by
 \begin{align}
     \sigma^2_R &= \frac{1+e^{-2r}}{4} \;, \quad \quad 
     \sigma^2_I = \frac{1+e^{2r}}{4} \;.
 \label{Het4}
 \end{align}
For displacement parameter estimation, it is generally assumed that the squeezing parameter ($r$) of the initial probe state is known \cite{morelli2021bayesian,wang2007quantum}. In the simplistic setup, the probe state does not undergo any squeezing which corresponds to $r=0$, while if one is interested to increase the estimation accuracy of one of the quadratures then that particular quadrature of the probe state can be squeezed by pumping a degenerate optical parametric amplifier (of known gain) with the probe state \cite{weedbrook2012gaussian,wang2007quantum}.

We focus our attention on estimating the prior parameters $\alpha_{0,R}, \sigma_{0,R}^2$ and $\alpha_R$ from the observed data. Note that similar analysis can be mirrored for the imaginary part $\alpha_{0,I}, \sigma_{0,I}^2$, $\alpha_I$. Optimal Bayesian estimation of an unknown parameter with Gaussian prior and Gaussian likelihood function is detailed in Appendix \ref{App_EM}. Using (\ref{app3})-(\ref{app5}) from Appendix \ref{App_EM}, the posterior distribution of  $\alpha_R$ is Gaussian with mean and variance given by
\begin{equation}
   \mu_{\alpha_R} = \left( \frac{1}{\sigma_{0,R}^2} + \frac{M}{\sigma_{R}^2} \right)^{-1} \left( \frac{\sum_{i=1}^{M}\beta_{R}^{i}}{\sigma_{R}^2 } + \frac{\alpha_{0,R}}{\sigma_{0,R}^2} \right) \;, 
   \label{Het7}
\end{equation}
\begin{equation}
  \sigma_{\alpha,R}^2 = \left( \frac{1}{\sigma_{0,R}^2} + \frac{M}{\sigma_{R}^2} \right)^{-1}  \;.
  \label{Het6}
\end{equation}
Therefore, the optimal Bayesian estimate of $\alpha_{R}$ is given by the posterior mean, i.e., $\hat{\alpha}_R=\mu_{\alpha_R}$. Note that for $M=1$, the result given by (\ref{Het7}) agrees with the result of \cite[Eq.~20]{morelli2021bayesian}. The prior parameters $\alpha_{0,R}, \sigma_{0,R}^2$ required in (\ref{Het7}) can be estimated by using an EM algorithm which is detailed in Appendix \ref{App_EM}, and summarized in Algorithm \ref{Algo1}. The estimated values $\Hat{\alpha}_{0,R}, \Hat{\sigma}_{0,R}^2$  obtained from Algorithm \ref{Algo1} can be substituted in (\ref{Het7}) to obtain the estimate of the real part of the displacement parameter $\alpha_{R}$. 
\begin{algorithm}
\DontPrintSemicolon
\SetAlgoLined
\KwInput{$\bm{\beta}_R, \sigma_R^2, M $}
\KwOutput{$\Hat{\alpha}_{0,R}, \Hat{\sigma}_{0,R}^2$ }
  Set $t=0$ \\
  Initialize  with \ $ \sigma_{0,R,t}^2 \sim U[0,1]$ and $\ \alpha_{0,R,t} \sim \mathcal{N}(0,1) $ \\
 \Repeat{\text{convergence}}
 {
  $\sigma_{0,R,t+1}^2= \frac{\sigma_R^2 \sigma_{0,R,t}^2}{\sigma_R^2 + M\sigma_{0,R,t}^2}$ \;
  $\alpha_{0,R,t+1} = \sigma_{0,R,t+1}^2 \left(\frac{\sum_{i=1}^M \beta_R^i}{\sigma_R^2} + \frac{\alpha_{0,R,t}}{\sigma_{0,R,t}^2} \right)$\;
  Set $t=t+1$\;
 }
 \caption{EM Algorithm for Heterodyne Measurement.}
 \label{Algo1}
\end{algorithm}

A similar method can be used for estimating the prior parameters of the imaginary part $\Hat{\alpha}_{0,I}, \Hat{\sigma}_{0,I}^2$ by replacing $\sigma_{R}^2, \bm{\beta}_R$ with $\sigma_{I}^2, \bm{\beta}_I $ respectively in Algorithm \ref{Algo1}. Subsequently, these estimated values can be substituted in (\ref{Het7}) to obtain $\Hat{\alpha}_{I}$.

\begin{figure}[htp] 
\centering
\subfigure[$\alpha_R$ ]{
\includegraphics[width=0.5\textwidth]{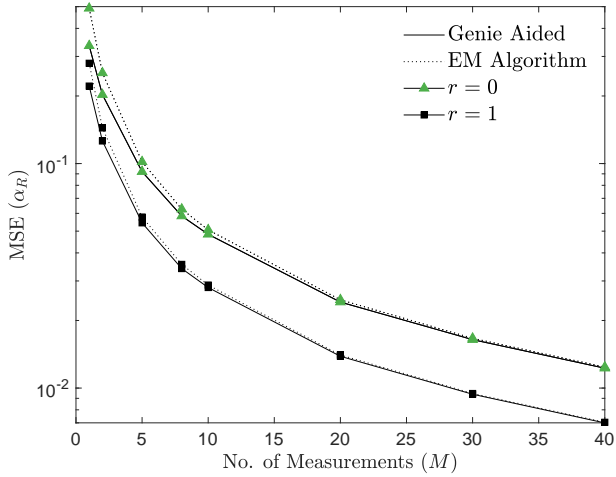}%
\label{figr1_a}
}\hfil
\subfigure[$\alpha_I$]{
\includegraphics[width=0.5\textwidth]{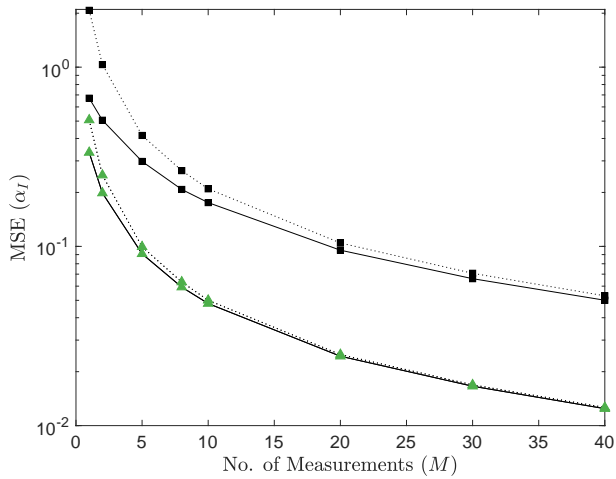}%
\label{figr1_b}
}
\caption{The plots show the MSE of $\alpha_R$ and $\alpha_I$ as the number of measurements $M$ increases for two different squeezing parameters $r=0,1$ with heterodyne measurement. Results are shown for the proposed EM algorithm and the `Genie Aided' Bayesian estimation method proposed in \cite{morelli2021bayesian} that required perfect knowledge of the prior distribution parameters. The other simulation parameters are $\alpha_{0,R}=\alpha_{0,I}=2$, $\sigma_{0,R}^2=\sigma_{0,I}^2=1$. }
\label{Fig1}
\end{figure}

We show the performance of the proposed EM algorithm for estimating $\alpha_R$ by considering a simulation scenario with $\alpha_{0,R}=\alpha_{0,I}=2$, $\sigma_{0,R}^2=\sigma_{0,I}^2=1$ and two different squeezing scenarios with $r=0$ and $r=1$. We also compare the performance with the previously proposed `Genie Aided' Bayesian method \cite{morelli2021bayesian}, that requires perfect knowledge of the prior distribution parameters. As a key difference, our proposed algorithm estimates these parameters from the measured data only. We use the mean squared error (MSE) as the metric for comparing the performance of the algorithms. The MSE metric is defined as
\begin{align}
    \text{ MSE}\left( \alpha_R \right) &= \mathbb{E}\left[|\alpha_R-\Hat{\alpha}_R|^2 \right], \nonumber \\
    \text{ MSE}\left( \alpha_I \right) &= \mathbb{E}\left[|\alpha_I-\Hat{\alpha}_I|^2 \right] \;.
    \label{em9}
\end{align}
Fig.\ \ref{Fig1} shows the MSE comparison for our proposed EM algorithm and the `Genie Aided' method that has perfect knowledge of prior distribution parameters, as the number of observations $M$ increases. The `Genie Aided' method is similar to the Bayesian estimation method proposed in \cite{morelli2021bayesian}, however here we consider the multiple measurement case. It can be observed that as $M$ increases the performance of our proposed EM algorithm converges to that of the Bayesian method of \cite{morelli2021bayesian}, which requires perfect knowledge of the prior distribution parameters. Thus, our proposed EM algorithm is able to estimate the prior distribution parameters from the observed data only and can efficiently estimate the displacement parameter $\alpha$ without any knowledge of the prior distribution parameters. Further, it can be observed that as the squeezing parameter $r$ increases the MSE of $\alpha_R$ decreases while the MSE of $\alpha_I$ increases. This observation is consistent with the theory of Gaussian quantum states, since from (\ref{Het4}), it can be observed that for $r>0$, $\sigma_R^2$ decreases and consequently the posterior variance of $\alpha_R$ in (\ref{Het6}) decreases. Similarly, $\sigma_I^2$ in (\ref{Het4}) decreases with increasing $r$, and consequently the posterior variance of $\alpha_I$ increases. Further, we note that for $r>0$, the variance of the observed data $\bm{\beta}_I$ increases, hence the EM algorithm requires more data (larger $M$) for the convergence of its performance to the `Genie Aided’ bound.

\subsection{Homodyne Measurement}
In this section we present the optimal Bayesian estimation of the displacement parameter, assuming a homodyne measurement. In homodyne measurement for measuring the quadrature $\hat{q}$, the POVM is given by $\{ \ket{q}\bra{q} \}_{q \in \mathbb{R}}$. Although the displacement $\alpha$ can be complex $\alpha=\Re[\alpha]+i\Im[\alpha]$, the measurement outcome from homodyne measurement with POVM $\{\ket{q} \bra{q} \}_{q \in \mathbb{R}}$ helps to estimate only the real part, i.e., $\alpha_{\mathrm{R}}=\Re[\alpha]$. A similar procedure can be used to estimate the imaginary part $\alpha_{\mathrm{I}}=\Im[\alpha]$ by introducing a $\pi/2$ phase shift to the local oscillator of the homodyne detector such that the orthogonal quadrature $\Hat{p}$ of the probe state is measured. As before, we consider a multiple measurement scenario, where $M$ measurements are made and the outcomes are i.i.d with the probability to observe $q_i$, given a displacement $\alpha_R$, for the $i$-th measurement given by
\begin{equation}
    p(q_i|\alpha_R) = \frac{\exp\left[-\frac{(q_i-\sqrt{2}\alpha_R)^2}{\cosh{2r}-\sinh{2r}} \right]}{\sqrt{\pi(\cosh{2r}-\sinh{2r})}}
    \label{hom1}
\end{equation}
where $r$ is the squeezing parameter. Letting $\bm{q} = [q_1,q_2,\ldots,q_M]^T$ be the vector containing all the measurements, the conditional likelihood is given by
\begin{equation}
    p(\bm{q}|\alpha_R) = \mathcal{N} \left( \sqrt{2}\alpha_R \bm{1}_M, \sigma_q^2 \bm{I}_M \right)
    \label{hom2}
\end{equation}
where the variance parameter $\sigma_q^2$ is given by
\begin{equation}
    \sigma_q^2= \frac{\cosh{2r}-\sinh{2r}}{2} \; .
    \label{hom3}
\end{equation}
Using (\ref{app3})-(\ref{app5}) from Appendix \ref{App_EM}, the posterior distribution of $\alpha_R$ is Gaussian with mean and variance given by
\begin{equation}
    \mu_{\alpha_R} = \left( \frac{1}{\sigma_{0,R}^2} + \frac{2M}{\sigma_{q}^2} \right)^{-1} \left( \frac{\sqrt{2}\sum_{i=1}^{M}q_i}{\sigma_{q}^2 } + \frac{\alpha_{0,R}}{\sigma_{0,R}^2} \right) \;,
     \label{hom6}
\end{equation}
and 
\begin{equation}
   \sigma^2_{\alpha_R}= \left( \frac{1}{\sigma_{0,R}^2} + \frac{2M}{\sigma_{q}^2} \right)^{-1} \;.
   \label{hom7}
\end{equation}
Thus, the optimal Bayesian estimate of $\alpha_R$ is given by $\hat{\alpha}_R = \mu_{\alpha_R}$. As before, the prior distribution parameters $\alpha_{0,R}$ and $\sigma_{0,R}^2$ can be estimated by using a similar EM algorithm as explained in Appendix \ref{App_EM}, and is summarized in Algorithm \ref{Algo2}. The estimated values $\Hat{\alpha}_{0,R}, \Hat{\sigma}_{0,R}^2$  obtained from Algorithm \ref{Algo2} can be substituted in (\ref{hom6}) to obtain the estimate of the real part of the displacement parameter $\alpha_{R}$. 
\begin{algorithm}
\DontPrintSemicolon
\SetAlgoLined
\KwInput{$\bm{q}, \sigma_q^2, M $}
\KwOutput{$\Hat{\alpha}_{0,R}, \Hat{\sigma}_{0,R}^2$ }
  Set $t=0$\\
  Initialize  with \ $ \sigma_{0,R,t}^2 \sim U[0,1]$ and $\ \alpha_{0,R,t} \sim \mathcal{N}(0,1) $ \\
 \Repeat{\text{ convergence}}
 {
  $\sigma_{0,R,t+1}^2= \frac{\sigma_q^2 \sigma_{0,R,t}^2}{\sigma_q^2 + 2M\sigma_{0,R,t}^2}$ \;
  $\alpha_{0,R,t+1} = \sigma_{0,R,t+1}^2 \left(\frac{\sqrt{2} \sum_{i=1}^M q_i}{\sigma_q^2} + \frac{\alpha_{0,R,t}}{\sigma_{0,R,t}^2} \right)$\;
  Set $t=t+1$\;
 }
 \caption{EM Algorithm for estimating $\alpha_{0,R}, \sigma_{0,R}^2$ }
 \label{Algo2}
\end{algorithm}

\begin{figure}[htp] 
\centering
\subfigure[$r=0$ ]{%
\includegraphics[width=0.5\textwidth]{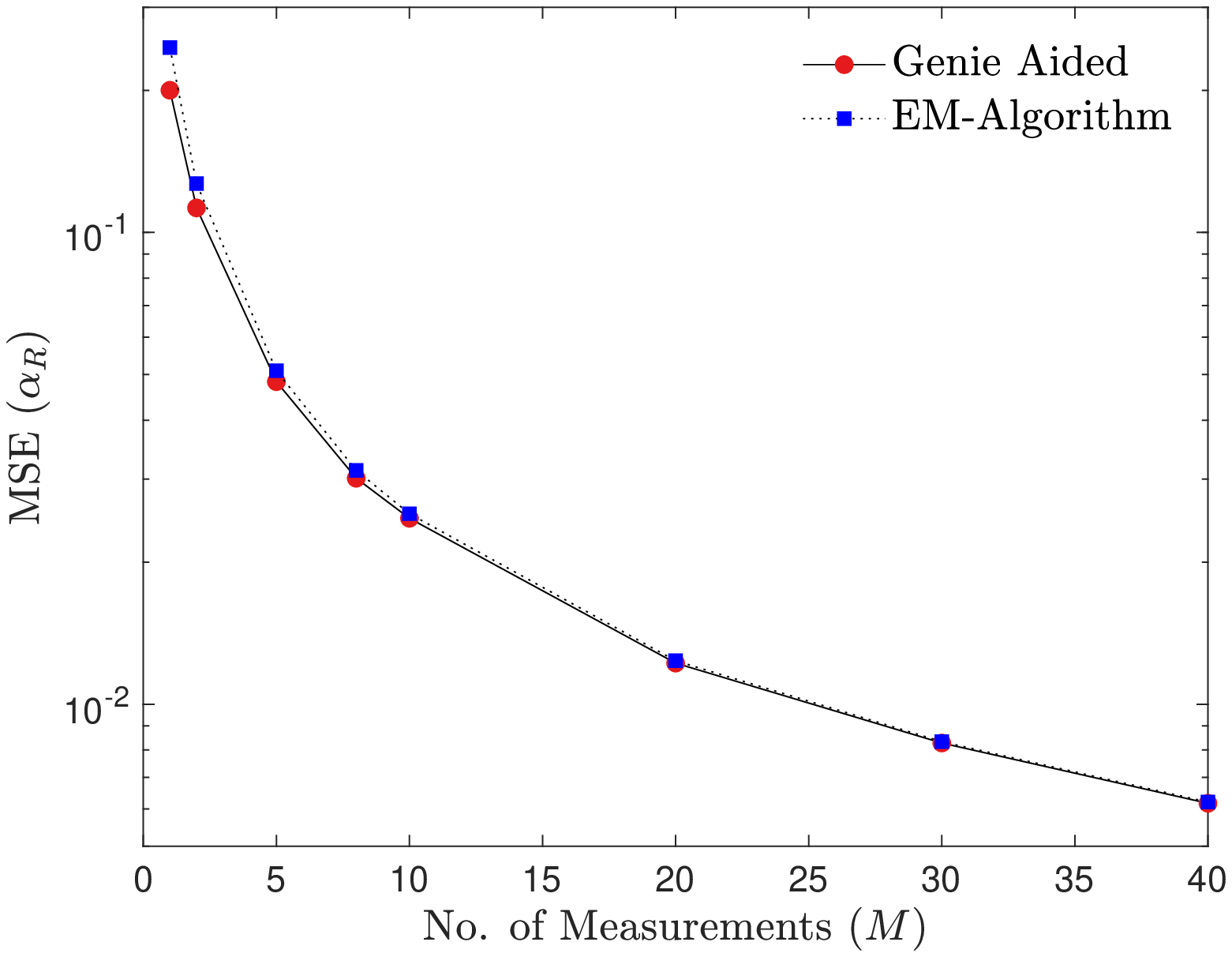}%
\label{figh1:a}%
}\hfil
\subfigure[$r=1$]{%
\includegraphics[width=0.5\textwidth]{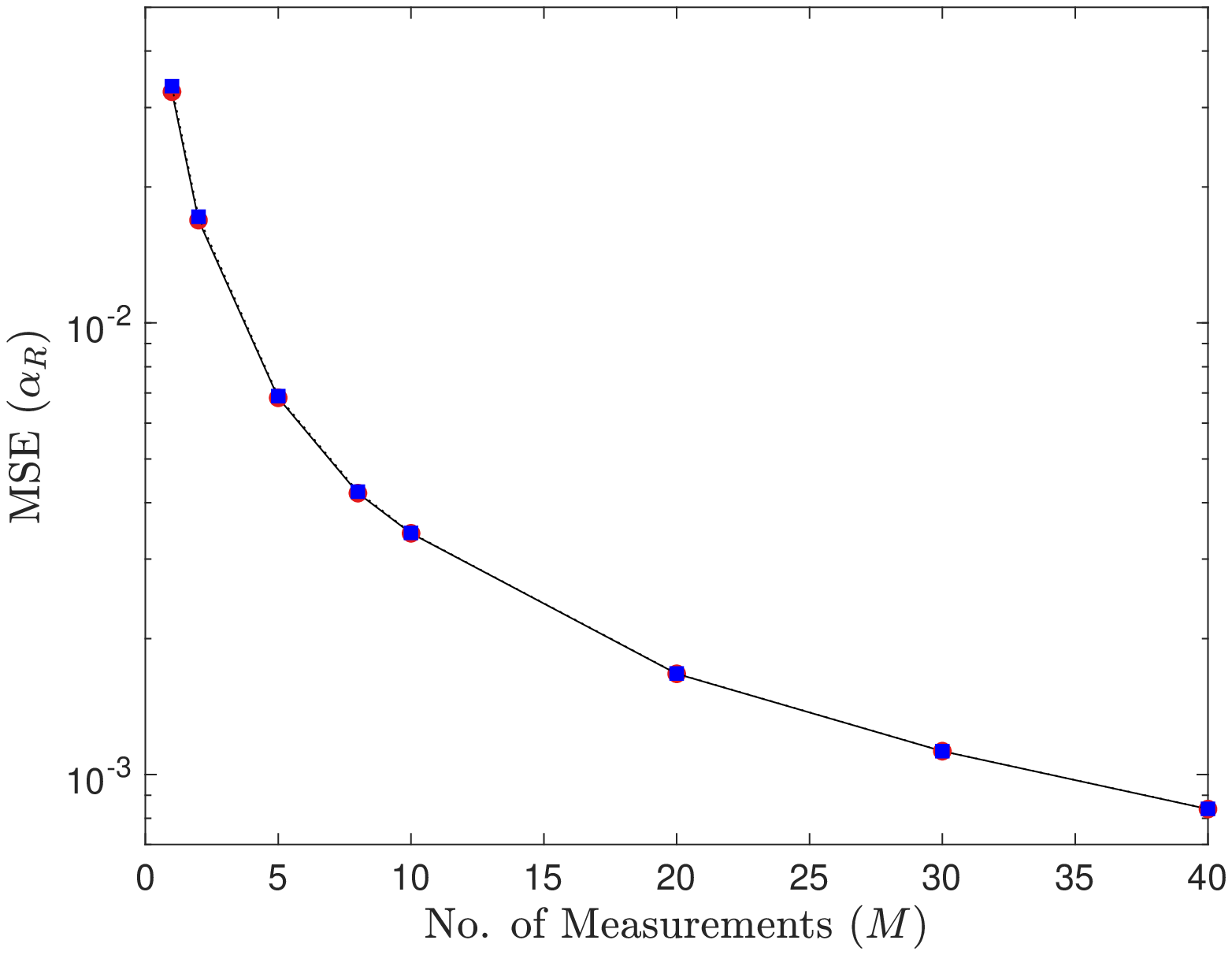}%
\label{figh1:b}%
}
\caption{ The plots show the MSE of $\alpha_R$ as the number of measurements $M$ increases for two different squeezing parameters (a) $r=0$, and (b) $r>0$ with homodyne measurement. Results are shown for the proposed EM algorithm and the `Genie Aided' Bayesian estimation method proposed in \cite{morelli2021bayesian} that required perfect knowledge of the prior distribution parameters. The other simulation parameters are $\alpha_{0,R}=2, \sigma_{0,R}^2=1$.}
\label{Fig_homo_real}
\end{figure}

We compare the performance of the proposed EM algorithm for estimating $\alpha_R$ by considering a simulation scenario similar to the heterodyne case. Fig.\ \ref{Fig_homo_real} shows the MSE comparison of the proposed EM algorithm and the previous algorithm of \cite{morelli2021bayesian} for two different squeezing parameters $r=0,1$. The simulation results reveal that the MSE performance of the proposed EM algorithm is very close to that of ‘Genie Aided’ Bayesian estimator of \cite{morelli2021bayesian}, that assumes perfect knowledge of the prior parameters. Further, we observe that for $r>0$, the EM algorithm achieves a faster convergence to the `Genie Aided' lower bound since the variance of the measurements in (\ref{hom1}) decreases for $r>0$. 


\section{Squeezing Estimation} \label{squeezing}
We now turn to the problem of squeezing parameter estimation of Gaussian quantum states. We consider the problem of estimating the squeezing strength $r \in \mathbb{R}$ of the operator $\hat{S}(r)$ defined in (\ref{sec2.7}), that acts on the coherent probe state $\ket{\alpha}$. In practical experiments, the problem of squeezing parameter estimation is very relevant where a degenerate parametric amplifier is pumped by a strong coherent field, and one is interested to find the optimal amplifier gain \cite{chiribella2006optimal}. We note that for squeezing parameter estimation, it is generally assumed that the displacement of the probe state is known which is related to the average number of photons (energy of the coherent field), and this is a practically feasible assumption \cite{monras2006optimal,chiribella2006optimal,morelli2021bayesian}.

Neither the heterodyne nor the homodyne measurement schemes are the optimal covariant measurement strategies for estimating the squeezing \cite{chiribella2006optimal,milburn1994hyperbolic}. The prior work \cite{morelli2021bayesian} considered a homodyne measurement strategy and proposed a sub-optimal numerical method for Bayesian estimation of the squeezing parameter. In this work, we first consider the optimal POVM scheme proposed in \cite{chiribella2006optimal} along with multiple measurements, and propose a machine learning based method for estimating $r$. In contrast to \cite{chiribella2006optimal}, that proposed a maximum likelihood (ML) estimate, we propose a Bayesian estimation scheme using the optimal POVM measurement outcomes. We also consider the case of suboptimal homodyne measurements, that have a practical advantage of easier implementation in laboratory. In this case, we present an ML estimate of $r$, since it is challenging to analytically characterize the corresponding Bayesian estimate.

\subsection{ Optimal POVM}
We consider a Gaussian prior for the squeezing parameter $r$,
\begin{equation}
    p(r) = \frac{1}{\sqrt{2\pi\sigma_0^2 }} \exp(-\frac{(r-r_0)^2}{\sigma_0^2}) \;
    \label{sq2} 
\end{equation}
with prior mean $r_0$ and variance $\sigma_0^2$.
We consider a multiple measurement scenario where $M$ independent measurements are made for estimating $r$. The optimal POVM for estimating $r$ is given by \cite[Eq.~12]{chiribella2006optimal}. The likelihood to observe $\xi_i$ in the $i$-th measurement is then given by \cite[Eq.~19]{chiribella2006optimal}
\begin{equation}
    p(\xi_i|r) = \sqrt{\frac{2|\alpha|^2}{\pi}} e^{-2|\alpha|^2(\xi_i-r)^2} \;.
    \label{sq3}
\end{equation}
Letting $\bm{\xi}=[\xi_1,\xi_2,\ldots,\xi_M]$ be the vector containing all the observations obtained from the POVM, the conditional likelihood of the observed data is given by
\begin{equation}
    p(\bm{\xi}|r)=\mathcal{N}\left(r\bm{1}_M, \sigma^2\bm{I}_M \right) \;,
    \label{sq4}
\end{equation}
where $\sigma^2= \frac{1}{4|\alpha|^2}$.
Similar to displacement estimation, we estimate the prior distribution parameters $r_0, \sigma_0^2$ and then find $\hat{r}$ from the observed data only. Using (\ref{app3})-(\ref{app5}) from Appendix \ref{App_EM}, the posterior distribution of $r$ is Gaussian with mean and variance
\begin{equation}
    \mu_{r} = \left( \frac{1}{\sigma_{0}^2} + \frac{M}{\sigma^2} \right)^{-1} \left( \frac{\sum_{i=1}^{M}\xi_i}{\sigma^2 } + \frac{r_{0}}{\sigma_{0}^2} \right) 
     \label{sq7}
\end{equation}
and
\begin{equation}
  \sigma_{r}^2=  \left( \frac{1}{\sigma_{0}^2} + \frac{M}{\sigma^2} \right)^{-1}  \;.
  \label{sq6}
\end{equation}
The point estimate of $r$ is given by the posterior mean, i.e., $\Hat{r} = \mu_{r}$. The prior parameters $r_{0}, \sigma_{0}^2$ required for evaluating $\mu_{r}$ can be estimated from the observed measurement $\bm{\xi}$ by using the EM algorithm presented in Appendix \ref{App_EM}. The estimated values $\Hat{r}_{0}, \Hat{\sigma}_{0}^2$ obtained from the EM algorithm can be substituted in (\ref{sq7}) to obtain the estimate of the squeezing parameter $r$.

We show the performance of our proposed EM algorithm with POVM measurement for estimating $r$. We consider a simulation scenario with $r_0=1\; \text{ and}\; \sigma_0^2=0.5$. Fig. \ref{Fig_sq1} shows the MSE of $r$ as the number of observations $M$ increases for two different values of $\alpha$. Results are shown for the optimal POVM measurement scheme with the proposed EM algorithm, along with the `Genie Aided' algorithm that uses perfect knowledge of the prior distribution parameters $r_0, \sigma_0^2$. It can be observed that the MSE performance of the proposed EM algorithm is very close to that of the `Genie Aided' bound. Further, we observe that the MSE decreases as the displacement of the initial probe state $|\alpha|$ increases since the variance of the observed data in (\ref{sq3}) is inversely proportional to $|\alpha|^2$.

\subsection{ Homodyne Measurement}
Next we consider the sub-optimal homodyne measurement scheme proposed in \cite{morelli2021bayesian} for squeezing parameter estimation. The authors of \cite{morelli2021bayesian} considered a single measurement scenario and numerically evaluated the posterior variance of the squeezing parameter without giving an explicit estimate of the squeezing parameter. The authors only presented an efficient estimate of the squeezing parameter for the case of vacuum probe state. Here, we present an efficient ML estimation scheme for the multiple measurement case with homodyne measurement for general coherent probe states. Note that here we do not use a heterodyne measurement scheme since the variance of the measurement outcome is higher for the heterodyne scheme, limiting the accuracy of squeezing estimation. Homodyne measurement is not a covariant measurement for squeezing estimation, but it is a Gaussian measurement which can be easily realized in practice. The likelihood to observe $q_i$ during in the $i$-th measurement is Gaussian \cite[Eq.~51]{morelli2021bayesian}
\begin{equation}
    p(q_i|r)= \frac{\exp\left[-e^{2r}(q_i-\sqrt{2}\alpha_R e^{-r})^2\right]}{e^{-r}\sqrt{\pi}} \;.
    \label{sq8}
\end{equation}
Therefore, the likelihood of the observed data $\bm{q}=[q_1,q_2,\ldots,q_M]$ is given by
\begin{equation}
    p(\bm{q}|r) = \mathcal{N}\left(\sqrt{2}\alpha_R e^{-r} \bm{1}_M, \frac{e^{-2r}}{2}\bm{I}_M \right) \;.
    \label{sq9}
\end{equation}
We note that the unknown parameter $r$ is embedded in both the mean and variance of the likelihood function through the $\exp$ function. Thus, it is difficult to find the posterior distribution of $r$. Further, the multiple measurement data makes it complicated to numerically evaluate the posterior mean and variance as done in \cite{morelli2021bayesian} for the single measurement case. Therefore, here we find the ML estimate of $r$. The ML estimate of $r$ is given by
\begin{align}
     \hat{r}_{\text{ ML}} & = \underset{r}{\text{argmax}}  \, \log p(\bm{q}|r) \;,
     \label{sq10}
\end{align}
where the log-likelihood function $\log p(\bm{q}|r)$ admits
\begin{align}
    \log p(\bm{q}|r) = -\frac{M}{2}\log\pi+Mr - e^{2r}\sum_{i=1}^{M}(q_i-\sqrt{2}\alpha_Re^{-r})^2 \,.
    \label{sq10a}
\end{align}
The ML estimate can be obtained by solving $f'(r)=0$, where $f(r)$ is defined as
\begin{align}
    f(r)= Mr - e^{2r}\sum_{i=1}^{M}\left(q_i^2+2\alpha_R^2 e^{-2r}-2\sqrt{2}\alpha_{R} e^{-r}q_i\right)\;.
    \label{sq10b}
\end{align}
After some algebra, we obtain the following equation
\begin{equation}
    2q''e^{2r}-2\sqrt{2}\alpha_R q' e^{r}-M=0 \;,
    \label{sq11}
\end{equation}
where $q'= \sum_{i=1}^{M}q_i$ and $q'' = \sum_{i=1}^{M}q_i^2$. The above non-linear equation in $r$ can be solved by a change of variable $t=e^{r}$, and then solving the resultant quadratic equation for $t$. Finally, after some algebraic manipulations we obtain
\begin{equation}
    \hat{r}_{\text{ ML}} = \log\left( \frac{2\sqrt{2} \alpha_Rq'+ \sqrt{8\alpha_R^2 q'^{2}+8Mq''}}{4q''} \right) \;.
    \label{sq12}
\end{equation}

\begin{figure}[ht] 
\centering
\subfigure[$\alpha=1+1i$ ]{%
\includegraphics[width=0.5\textwidth]{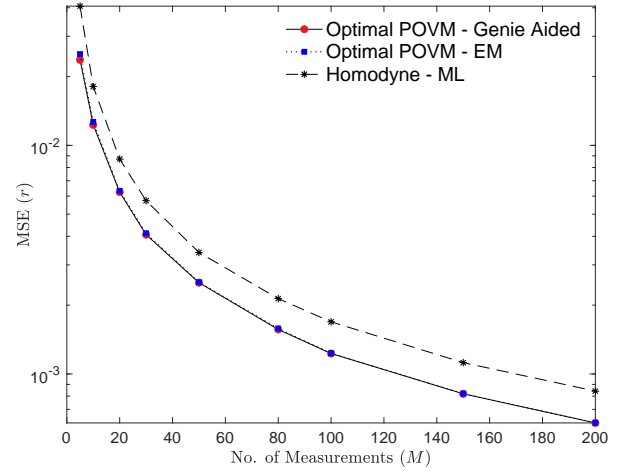}%
\label{Fig_sq1:a}%
}\hfil
\subfigure[$\alpha=2+2i$]{%
\includegraphics[width=0.5\textwidth]{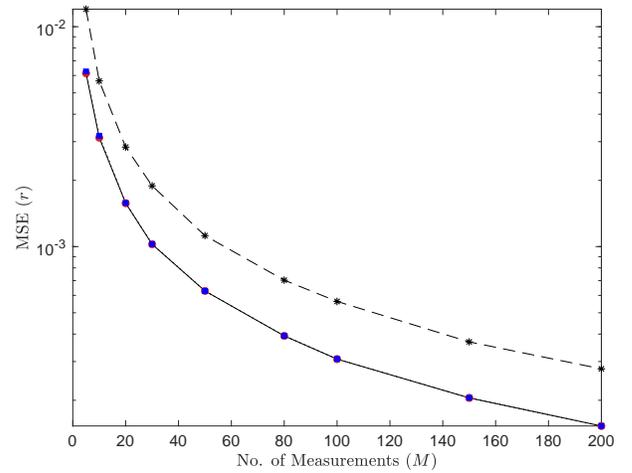}%
\label{Fig_sq1:b}%
}
\caption{ The plots show the MSE of the squeezing parameter $r$ as the number of measurements $M$ increases for two different values of displacement parameter $\alpha$. Results are shown for the proposed EM based estimate obtained from the optimal POVM measurement \cite{chiribella2006optimal}, and the ML estimate obtained from the suboptimal Gaussian homodyne measurement scheme proposed in \cite{morelli2021bayesian}. The `Genie Aided' lower bound corresponds to the case when the prior distribution parameters $r_0, \sigma_0^2$ are perfectly known. }
\label{Fig_sq1}
\end{figure}

\begin{figure}[htp] 
\centering{%
\includegraphics[width=0.5\textwidth]{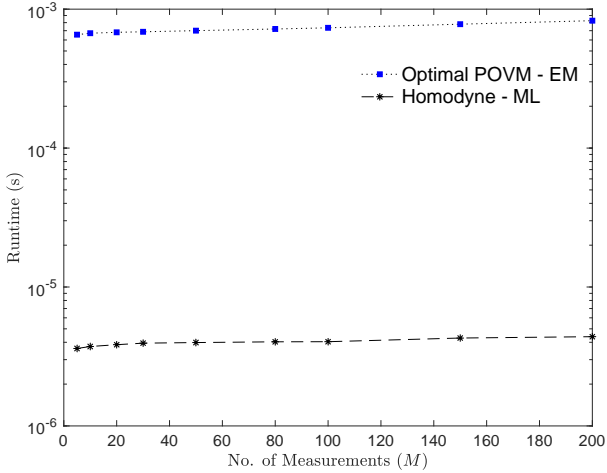}%
}
\caption{The plots compare the runtime of two estimation algorithms for squeezing estimation. Results are shown for the proposed EM based estimate obtained from the optimal POVM measurement and the suboptimal ML estimate with homodyne measurement. }
\label{Fig_sq_runtime}
\end{figure}


Fig. \ref{Fig_sq1} compares the MSE performance of the proposed ML estimate (with homodyne measurement) of $r$ against the corresponding EM-based estimate (with POVM measurement) derived previously. It is observed that the optimal POVM measurement scheme has a lower MSE than the homodyne measurement scheme. Similar to the POVM case, the MSE of the ML estimate with homodyne measurement decreases as the displacement of the initial probe state $|\alpha|$ increases. Moreover, the performance gap between the optimal POVM measurement and homodyne measurement increases as $|\alpha|$ increases.


In terms of computational complexity, the ML estimate has a lower complexity since it has a closed form expression, whereas the complexity of the optimal POVM is higher due to the iterative EM algorithm. Fig.\ \ref{Fig_sq_runtime} compares the runtime of the two estimation algorithms for squeezing estimation. As expected, we observe that the EM algorithm has a higher runtime that can be around two orders of magnitude higher than that of the ML estimate.

In a practical setting, the choice of the optimal POVM or the homodyne measurement scheme depends on the availability of specific equipment in the laboratory. In general, homodyne measurement is easier to implement, making it a popular choice among experimentalists \cite{morelli2021bayesian}. However, if a higher accuracy is desired then the optimal POVM need to be implemented, requiring additional equipment with higher costs \cite{chiribella2006optimal}. 


\section{Phase Estimation} \label{phase}
In this section we study the problem of phase estimation for Gaussian quantum states. We consider the problem of estimating the parameter of an unknown phase rotation operator acting on a coherent Gaussian probe state. Similar to the displacement estimation scenario, we assume that multiple independent measurements are obtained and then our proposed machine learning based method is used on the observed data to obtain the phase estimate. Unlike the previous work \cite{morelli2021bayesian} that considered a uniform prior for the phase parameter, in this work we consider a von Mises prior that is widely used for modelling directional parameters \cite{quinn2011bayesian,das2011variational,barakat1988probability,khatri1977mises,wang2012performance,nielsen2011bayesian}. We assume a coherent probe state $\ket{\alpha}$, and, without loss of generality, that the displacement is positive and real, i.e., $\alpha=|\alpha|>0$. This probe state is acted upon by the rotation operator $\hat{R}(\theta)$ given by (\ref{sec2.8}). The action of the rotation operator $\hat{R}(\theta) $ on the probe state $\ket{\alpha}$ gives the encoded state $\ket{e^{-i\theta} \alpha}$. Similar to the squeezing estimation problem, it is generally assumed that the displacement parameter of the initial probe state is known, and the problem is to estimate the unknown phase parameter $\theta$ \cite{monras2006optimal,morelli2021bayesian}.

Since heterodyne detection measures both the quadratures it can estimate the phase in the entire interval $\theta \in [-\pi, \pi)$. On the other hand, the homodyne measurement can only estimate the phase in the interval $\theta \in [0, \pi]$, since it cannot distinguish between phases $\theta$ and $-\theta$ owing to its ability to measure only one of the quadratures. Therefore, in this paper we focus only on the heterodyne measurement scheme for phase estimation. We propose an empirical Bayes method to estimate the prior distribution parameters as well as the optimal Bayes estimate of $\theta$ using the observed data from heterodyne measurement.

As before, we consider a multiple measurement scenario where $M$ independent measurements are made. The outcomes are i.i.d with the likelihood to observe $\beta_i \in \mathbb{C}$, given that the phase shift is $\theta$, given by \cite[Eq.~33]{morelli2021bayesian}
\begin{equation}
    p(\beta_i|\theta) = \frac{1}{\pi}|\braket{\beta_i}{e^{-i\theta}\alpha}|^2 = \frac{1}{\pi}e^{-|e^{i\theta}\beta_i-\alpha|^2} \;.
    \label{ph2}
\end{equation}
Letting $\bm{\beta}=[\beta_1,\beta_2,\ldots,\beta_M]^T $ be the vector containing the measurement outcomes, the likelihood of the observed data is given by
\begin{equation}
    p\left(\bm{\beta}|\theta \right) = \mathcal{CN}\left(e^{-i\theta}\alpha \bm{1}_M, \bm{I}_M \right) \;.
    \label{ph_vm1}
\end{equation}
We assume a von Mises prior on $\theta$, which is a circular distribution that has been widely used in classical signal processing applications to model directional parameters \cite{quinn2011bayesian,das2011variational,barakat1988probability,khatri1977mises,wang2012performance,nielsen2011bayesian}. In quantum phase estimation, a recent work assumed a wrapped Gaussian prior for qubit phase estimation \cite{friis2017flexible}. The wrapped Gaussian distribution is closely approximated by the von Mises distribution, but is less analytically tractable \cite{collett1981discriminating,pewsey2005discrimination}. Moreover, the choice of von Mises prior is supported by the fact that it is a conjugate prior for a Gaussian measurement data model \cite{quinn2011bayesian}, which is the case considered in this work for phase estimation with heterodyne measurement. The von Mises distribution is parameterized by a shaping parameter $\kappa_0 \in \mathbb{C}$, and has pdf \cite{quinn2011bayesian}
\begin{equation}
    p(\theta|\kappa_0) = \frac{1}{2\pi I_{0}\left(|\kappa_0| \right)}\exp\left[\text{ Re}\left(\kappa_0 e^{-i\theta} \right) \right]
    \label{ph_vm2}
\end{equation}
where $I_0(\cdot)$ is the zeroth order modified Bessel function of the first kind. The mean of this distribution is $ \mathbb{E}[\theta]= \angle{\kappa_0}$. Apart from appropriately modeling the phase parameter, it has also been previously shown that von Mises is a conjugate prior for a wide class of Gaussian observation models \cite{quinn2011bayesian}. The following proposition summarizes the conjugate prior result for a von Mises prior with Gaussian measurement model.

Proposition 1: Let the prior distribution of the unknown phase parameter $\theta$ be a von Mises distribution with parameter $\kappa_0$, such that the pdf is given by (\ref{ph_vm2}). If the likelihood of the observed data $\bm{y} \in \mathbb{C}^{M\times 1}$ is Gaussian, i.e., $p(\bm{y}|\theta)= \mathcal{CN}\left(e^{-i\theta} \bm{x}, \sigma^2 \bm{I}_M \right)$, then the posterior distribution of $\theta$ is also von Mises with parameter $\kappa_p = \kappa_0 + \frac{2}{\sigma^2}\bm{y}^H\bm{x}$ \cite{quinn2011bayesian}.
\begin{IEEEproof}
Follows directly from \cite[Eq.~(7)-(9)]{quinn2011bayesian}.
\end{IEEEproof}


Using Proposition 1 along with (\ref{ph_vm1}),(\ref{ph_vm2}), the posterior distribution of $\theta$ is given by \cite{quinn2011bayesian}
\begin{equation}
    p(\theta|\bm{\beta}) = \frac{1}{2\pi I_{0}\left(|\kappa_p| \right)}\exp\left[\text{ Re}\left(\kappa_p e^{-i\theta} \right) \right]
    \label{ph_vm3}
\end{equation}
where
\begin{equation}
    \kappa_p=\kappa_0+2 \alpha \Tilde{\beta}^* \;,
    \label{ph_vm4}
\end{equation}
with $\Tilde{\beta}^*= \sum_{i=1}^{M}\beta_i^*$. Therefore, the optimal Bayes estimate of the phase parameter is given by
\begin{equation}
   \hat{\theta} = \angle{\kappa_p} \;.
   \label{ph_vm5}
\end{equation}
We estimate the prior distribution parameter $\kappa_0$ from the observed data by using an empirical Bayes method. The empirical Bayes estimate of $\kappa_0$ is given by \cite{bishop2006pattern}
\begin{align}
    \hat{\kappa}_0 &= \underset{\kappa_0}{\text{ argmax}}  \; \log p(\bm{\beta}|\kappa_0) \nonumber \\
    &= \underset{\kappa_0}{\text{ argmax}}  \; \log  \frac{p(\bm{\beta}|\theta ; \kappa_0) p(\theta|\kappa_0)}{p(\theta|\bm{\beta};\kappa_0)}
    \label{ph_vm6}
\end{align}
where the second equality follows from Bayes theorem,
\begin{equation}
    p(\theta|\bm{\beta};\kappa_0)  = \frac{p(\bm{\beta}|\theta;\kappa_0) p(\theta|\kappa_0)}{p(\bm{\beta}|\kappa_0)} \;.
\end{equation}
Using (\ref{ph_vm1})-(\ref{ph_vm3}) in (\ref{ph_vm6}), and after some simplifications we obtain
\begin{equation}
    \hat{\kappa}_0 = \underset{\kappa_0}{\text{ argmax}} \log \frac{I_0(|\kappa_0+ 2\alpha \Tilde{\beta}^*| )}{I_0\left(|\kappa_0| \right)} \;.
    \label{ph_vm7}
\end{equation}
The optimization problem in (\ref{ph_vm7}) can be solved numerically to find $\hat{\kappa}_0$ which can then be used in (\ref{ph_vm4}), (\ref{ph_vm5}) to obtain $\hat{\theta}$. The overall algorithm for phase estimation is summarized in Algorithm \ref{Algo_phase}.

\begin{algorithm}
\DontPrintSemicolon
\SetAlgoLined
\KwInput{$\bm{\beta}, \alpha, M $}
\KwOutput{$\Hat{\theta}$ }
  Evaluate $\Tilde{\beta}^*= \sum_{i=1}^{M}\beta_i^*$ \\
  Numerically solve: $\hat{\kappa}_0 = \underset{\kappa_0}{\text{argmax}} \log \frac{I_0(|\kappa_0+ 2\alpha \Tilde{\beta}^*| )}{I_0\left(|\kappa_0| \right)}$ \\
  Set $\kappa_p=\hat{\kappa}_0 +2 \alpha \Tilde{\beta}^*$ \\
  Return $\hat{\theta} = \angle{\kappa_p}$
 \caption{Empirical Bayes Algorithm for Phase Estimation}
 \label{Algo_phase}
\end{algorithm}

\begin{figure}[htp] 
\centering{%
\includegraphics[width=0.5\textwidth]{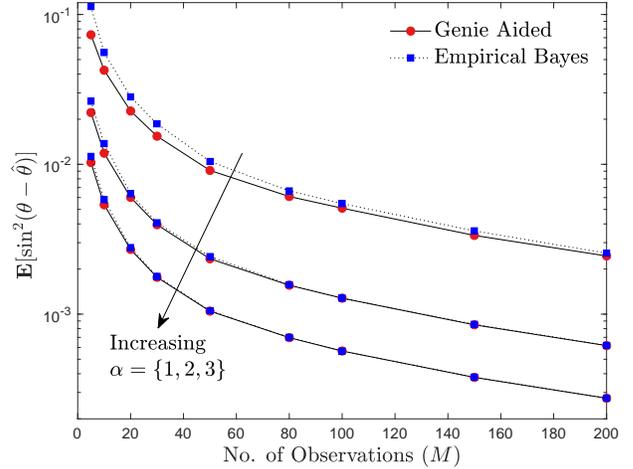}%
}
\caption{ The plot shows the performance of the proposed empirical Bayes method and the `Genie Aided' Bayes estimation scheme for phase estimation using heterodyne measurement as the number of measurements $M$ increases. Results are shown for $|\kappa_0|=4, \angle{\kappa_0}=0.5$ and different values of $\alpha=\{1,2,3\}$. }
\label{Fig_ph1}
\end{figure}

We evaluate the performance of the proposed empirical Bayes solution for phase estimation. Since the phase shifts are invariant under shifts by $2\pi$, we consider $\mathbb{E}[\sin^2(\theta-\Hat{\theta}) ]$ as the performance metric instead of the MSE metric considered before for displacement and squeezing estimation. Fig.\ \ref{Fig_ph1} shows the performance of our proposed empirical Bayes method and the `Genie Aided' Bayes estimation scheme as $M$ increases for different values of the initial probe state displacement $\alpha$. The `Genie Aided' estimation scheme requires perfect knowledge of the prior parameter $\kappa_0$, whereas the empirical Bayes method learns the prior parameter from the observed data. It is observed that the proposed empirical Bayes method converges to that of the `Genie Aided' bound as $M$ increases. Further, we observe that the estimation performance improves and the empirical Bayes method achieves a faster convergence to the `Genie Aided' bound as the displacement of the initial probe state $\alpha$ increases.  

\section{Conclusion} \label{conclusion}
We have proposed machine learning based methods for parameter estimation of continuous variable Gaussian quantum states. We focused on Bayesian estimation of the phase-space displacement, squeezing, and phase parameter of a single mode Gaussian quantum state. We considered a multiple measurement scenario and assumed a Gaussian prior on the displacement and squeezing parameter, and proposed an EM algorithm to estimate the prior distribution parameters from the measured data. For the phase estimation, we assumed a von Mises prior on the phase parameter and by considering a heterodyne measurement scheme we proposed an empirical Bayes method to estimate the prior distribution parameters from the observed data. The estimated prior parameters along with the observed data were used to find the optimum Bayesian estimate of the displacement, squeezing, and phase parameter. Our simulation results show that the proposed algorithms have estimation performance that is very close to that of ‘Genie Aided’ Bayesian estimators, that assume perfect knowledge of the prior parameters. In conclusion, the framework presented in this paper can be used by experimental physicists to find optimal estimates of quantum parameters using the measurement outcomes of their experiments.

In this work we focused on machine learning based single parameter estimation of single-mode Gaussian quantum states only. An interesting future work would be to extend the machine learning framework presented here for multi-parameter estimation of single and multi-mode Gaussian quantum states. This could include for example joint displacement and phase estimation, and joint squeezing and displacement estimation of single and multi-mode Gaussian quantum states.

\section{Appendix} \label{Appendix}
\subsection{Expectation Maximization (EM) Algorithm for Estimating Prior Distribution Parameters} \label{App_EM}

In this section we present the general EM algorithm for estimating the prior distribution parameters for a Gaussian observation model with Gaussian prior. Let the prior distribution of the unknown parameter $u$ be a Gaussian distribution with mean $\mu_0$ and variance $\sigma_0^2$, i.e.,
\begin{equation}
    p(u) = \frac{1}{\sqrt{2\pi\sigma_0^2}} \exp\left(-\frac{(u-\mu_0)^2}{2\sigma_0^2} \right)
    \label{app1} \;.
\end{equation}
Further, the likelihood of the observed data $\bm{y} \in \mathbb{R}^{M\times1}$ is Gaussian with
\begin{equation}
    p(\bm{y}|u) = \mathcal{N}\left(u\bm{g},\sigma_n^2 \bm{I}_M \right) \;.
 \label{app2}    
\end{equation}
Using properties of the Gaussian distribution, the posterior distribution of $u$ is given by \cite[Eq.~2.113-2.117]{bishop2006pattern}
\begin{equation}
    p(u|\bm{y}) = \mathcal{N} \left( \mu_{p}, \sigma_{p}^2 \right)
    \label{app3}
\end{equation}
where 
\begin{equation}
  \sigma_{p}^2=  \left( \frac{1}{\sigma_{0}^2} + \frac{\bm{g}^T\bm{g}}{\sigma_{n}^2} \right)^{-1}  \;,
   \label{app4}
\end{equation}
and
\begin{equation}
    \mu_{p} = \left( \frac{1}{\sigma_{0}^2} + \frac{\bm{g}^T\bm{g}}{\sigma_{n}^2} \right)^{-1} \left( \frac{\bm{g}^T \bm{y}}{\sigma_{n}^2 } + \frac{\mu_{0}}{\sigma_{0}^2} \right) \;.
     \label{app5}
\end{equation}
The prior distribution parameters $\bm{\theta} = [\mu_0,\sigma_0^2]$ can be estimated by maximizing the log-likelihood of the observed data, i.e., by solving the following optimization problem
\begin{equation}
    \left(\hat{\mu}_0,\hat{\sigma}_0^2 \right) = \text{argmax} \log p\left( \bm{y}|\bm{\theta}\right) \;.
     \label{app6}
\end{equation}
It is difficult to find the prior distribution parameters $\bm{\theta}$ by directly maximizing the log-likelihood of the observed measurements due to the constraint $\hat{\sigma}_{0}^2>0$ and the complexity of the objective function. Hence, we use the EM algorithm that maximizes a lower bound of the log-likelihood function. Note that $u$ is the latent variable that is not directly observed. The EM algorithm is an iterative algorithm that has two main steps. In the first step (E-step) of the $t$-th iterate, the expectation of the complete log-likelihood with respect to the posterior distribution of the latent variable is evaluated, and is denoted by $Q(\bm{\theta},\bm{\theta}_{t})$. In the second step (M-step), the parameters are updated $(\bm{\theta}_{t+1})$ by maximizing $Q(\bm{\theta},\bm{\theta}_{t})$. Let $ L(\bm{\theta}) = \log p\left( \bm{y}|\bm{\theta}\right)$ be the log-likelihood of the observed data, then $ L(\bm{\theta})$ is lower bounded as \cite{dempster1977maximum,sun2016majorization}
\begin{equation}
    L(\bm{\theta}) \geq Q(\bm{\theta},\bm{\theta}_{t})
    \label{app7}
\end{equation}
where 
\begin{equation}
   Q(\bm{\theta},\bm{\theta}_{t}) = \mathbb{E}_{p\left(u|\bm{y},\bm{\theta}_{t}  \right)}\left[ \log p\left(\bm{y},u|\bm{\theta} \right)\right] \;.
    \label{app8}
\end{equation}
From (\ref{app1}),(\ref{app2}) we obtain
\begin{align}
  \log p\left(\bm{y},u|\bm{\theta} \right) &= -\frac{1}{2} \Bigg(M\log2\pi\sigma_n^2 + \sum_{i=1}^{M} \frac{(y_i-ug_i)^2}{\sigma_n^2} \nonumber \\
  &\quad \quad \quad \quad + \log2\pi\sigma_{0}^2+ \frac{(u-\mu_{0})^2}{\sigma_{0}^2} \Bigg)  \;.
   \label{app9}
\end{align}
The posterior distribution of $u$ is given by (\ref{app3}). For the E-step, using (\ref{app3}) in (\ref{app8}) we obtain
\begin{align}
    Q(\bm{\theta},\bm{\theta}_{t}) &= -\frac{1}{2} \Bigg(M\log2\pi\sigma_n^2  + \sum_{i=1}^{M} \frac{(y_i-g_i \mu_{p})^2+ g_i^2\sigma_{p}^2}{\sigma_{n}^2} \nonumber \\
  &\quad  \; + \log2\pi\sigma_{0}^2  + \frac{(\mu_{0}-\mu_{p})^2+\sigma_{p}^2}{\sigma_{0}^2} \Bigg) \, ,
     \label{app10}
\end{align}
where we have used
\begin{align}
    \mathbb{E}_{p\left(u|\bm{y},\bm{\theta}_{t}  \right)} \left[ u \right] &= \mu_{p}\;, \nonumber \\
    \mathbb{E}_{p\left(u|\bm{y},\bm{\theta}_{t}  \right)} \left[ u^2 \right] &= \mu_{p}^2 + \sigma_{p}^2 \;.  \label{app10b}
\end{align}
For the M-step, we need to solve
\begin{equation}
     \frac{\partial Q(\bm{\theta},\bm{\theta}_{t})}{\partial \mu_{0}} = 0 \;, \quad \frac{\partial Q(\bm{\theta},\bm{\theta}_{t})}{\partial \sigma_{0}^2} = 0
    \label{app11}
\end{equation}
which give the updates
\begin{equation}
    \mu_{0,t+1} = \mu_{p} \;, \quad \sigma_{0,t+1}^2 = \sigma_{p}^2 \;. 
    \label{app12}
\end{equation}
The overall EM algorithm is summarized in Algorithm \ref{AlgoEM}. Note that the loop in Algorithm \ref{AlgoEM} terminates when the difference of $Q(\bm{\theta},\bm{\theta}_{t})$ in two consecutive iterations drops below a certain threshold, say $\epsilon = 10^{-3}$. The estimated values $\Hat{\mu}_{0}, \Hat{\sigma}_{0}^2$  obtained from Algorithm \ref{AlgoEM} can be substituted in (\ref{app5}) to obtain the optimal Bayesian estimate of $u$. 
\begin{algorithm}
\DontPrintSemicolon
\SetAlgoLined
\KwInput{$\bm{y}, \sigma_n^2, M $}
\KwOutput{$\Hat{\mu}_{0}, \Hat{\sigma}_{0}^2$ }
  Set $t=0$\\
  Initialize  with \ $ \sigma_{0,t}^2 \sim U[0,1]$ and $\ \mu_{0,t} \sim \mathcal{N}(0,1) $ \\
 \Repeat{\text{ convergence}}
 {
  $\sigma_{0,t+1}^2= \frac{\sigma_n^2 \sigma_{0,t}^2}{\sigma_n^2 + \bm{g}^T\bm{g} \sigma_{0,R,t}^2}$ \;
  $\mu_{0,t+1} = \sigma_{0,t+1}^2 \left(\frac{\bm{g}^T\bm{y}}{\sigma_n^2} + \frac{\mu_{0,t}}{\sigma_{0,t}^2} \right)$\;
  Set $t=t+1$\;
 }
 \caption{EM Algorithm for solving (\ref{app6})}
 \label{AlgoEM}
\end{algorithm}



\section*{Acknowledgment}
The authors thank Ayaka Usui for the insightful comments on the initial version of the draft that helped to further improve the content of the paper.

\bibliographystyle{IEEEtran}
\bibliography{IEEEabrv,tqe}

\end{document}